\documentstyle[12pt,fleqn]{article}

\def\beqa{\begin{eqnarray}}
\def\eeqa{\end{eqnarray}}
\def\beq{\begin{equation}}
\def\eeq{\end{equation}}

\def\vol{\int d^4x\,\sqrt{-g}}

\def\half{\frac{1}{2}}

\def\ln{{\rm ~ln}}

\def\umunu{^{\mu\nu}}
\def\dmunu{_{\mu\nu}}

\def\ddemu{_{;\mu}}  \def\udemu{^{;\mu}}
\def\ddenu{_{;\nu}}  
\def\ddea{_{;\alpha}}  \def\udea{^{;\alpha}}

\def\bib#1{$^{\ref{#1}}$}
\let\lam=\lambda  

\let\gam=\gamma

\let\alp=\alpha

\let\ome=\omega

\def\pr{{\it Phys. Rev.}\ }
\def\prl{{\it Phys. Rev. Lett.}\ }
\def\pl{{\it Phys. Lett.}\ }

\def\ijmp{{\it Int. J. Mod. Phys.}\ }

\def\nat{{\it Nature}\ }

\def\etal{{\it et al.}\ }
\def\ie{{\it i.e. }\ }

\def\rpr{ Phys. Rev. }
\def\rprl{ Phys. Rev. Lett. }
\def\rpl{ Phys. Lett. }
\def\rnp{ Nucl. Phys. }

\def\rcqg{ Class. Quantum Gravit. }

\def\cl{_{cl}}
\begin{document}
\null
\begin{flushright}
OAR 93/4\\
gr-qc/9303023
\end{flushright}
\centerline{\large \bf Inflationary attractors }
\centerline{\large \bf and perturbation spectra }
\centerline{\large\bf in generally coupled gravity}
\vspace{.3in}
\centerline{Luca Amendola, Diego Bellisai and Franco Occhionero}
\vspace{0.3in}
\centerline{{\it Osservatorio Astronomico di Roma}}
\centerline{\it Viale del Parco Mellini, 84}
\centerline{\it I-00136 Rome, Italy}
\vspace{.5in}
\baselineskip 14pt
\begin{quote}
A generic outcome of theories with scalar-tensor coupling is
the existence of inflationary attractors, either power-law
or de Sitter. The fluctuations arising during this phase are
Gaussian and their spectrum depends on the wavenumber $k$ according
to the power-law $k^{1/(1-p)}$, where $p$ is the
inflationary power-law exponent. We investigate to which extent these
properties depend on the coupling function and on the potential. We
find the class of models in which viable attractors exist. Within
this class, we find that the cosmic expansion and
the scaling of the fluctuation spectrum are independent
of the coupling function.
Further, the analytical solution of the
Fokker-Planck equation shows that the deviations from Gaussianity
are negligible.
\end{quote}
\vspace{.3in}
\centerline{Accepted Feb. 24, 1993}
\centerline{To be published on {\it Physical Review} D.}
\vfill

\vspace{.3in}
\baselineskip 14pt
\section{\normalsize\bf Introduction}
Non-minimally coupled theories of gravity are models in which a direct
coupling of the form
\beq\label{act1}
A_{NMC}=\vol \left[\half\xi f(\phi)R\right],
\eeq
between a matter field $\phi$ and the curvature scalar $R$ is added
to the gravity Action. The interesting history of such theories in
general relativity and
cosmology dates back to the Dirac suggestion
 and to Jordan-Brans-Dicke (JBD) models\bib{JBD}.
Terms with non-minimal coupling (NMC) arise quantizing
fields in curved space\bib{BD}, in multi-dimensional theories\bib{SS}
like superstrings or supersimmetry and in induced gravity theories\bib{ZA}.
On the other hand, La and Steinhardt's extended inflation\bib{LS} (EI)  and
many variants within the context of `first-order inflation'
\bib{AT} reconcile
a first-order phase transition with inflation
modulating  the bubble nucleation rate.
The cosmological kinematics of NMC theories is also discussed in
Ref. (\ref{KFM}).

No known fundamental principle predicts the functional form of
$f(\phi)$. One can find in literature
many {\it ans\"atze}: $f(\phi)=\phi^2$, as in the induced gravity model and
some version
of EI; $f(\phi)=a+b\phi^2+c\phi^4+...$, as in the hyperextended\bib{SA} design
of EI; $f(\phi)=e^{D\phi}$, as in dimensionally reduced Kaluza-Klein theories,
and so on. In Ref. [\ref{AL}], non-minimal derivative
couplings in $f(\phi)$ have also been taken into account.
Models with a generalized kinetic term like
$\omega(\phi)\phi\ddemu\phi\udemu/\phi$ and a coupling term
$\Phi R$ have also been considered\bib{SA} (the
case $\omega=const$ is indeed the JBD theory and the original formulation
of EI); up
to a redefinition of $\phi$ these models are equivalent to introducing
a coupling like (\ref{act1}). Aim
of this paper is to investigate in a systematic fashion the dynamical features
of models with a general coupling $f(\phi)$, without confining ourselves
to a specific choice, except for the following requests: that in the
large $\phi$ limit
the function  $f(\phi)$ grows faster than $\phi^2$ (the case
$f(\phi)\sim \phi^2$ being already well-known), and that the potential
$V(\phi)$, at least for large $\phi$,
can be written as
\beq\label{hyp}
V(\phi)=\lam f^M(\phi)\,,
\eeq
where $\lam$ and $M$ are arbitrary non-negative constants.
\footnote{
After this work was completed we learned of independent work
of De Ritis and coworkers\bib{DR}, who obtained exact solutions
in particular cases of the above functional relation.
}
We also assume $V(\phi),f(\phi)\ge 0$,
the equality being verified  when $\phi$ eventually
falls into its stable minimum.
It is remarkable that,
with the above  assumptions, it is possible to find analytically,
without specifying nor the coupling nor
the potential, the class of inflationary attractors,
the order-of-magnitude amplitude of the primordial perturbations
and the solution of the Fokker-Planck equation for
the stochastic fluctuations. Since $f(\phi)$ is left largely unspecified, we
speak here of theories with {\it general coupling }, instead of
non-minimal coupling models.
We also give examples of  numerical
phase-space (PS) of the investigated models.

The plausibility of  inflation depends on how large
is the set of initial conditions which evolve to inflationary
expansion of the scale factor: the existence of an inflationary
attractor extending to values of the field as large as
possible, up to the Planck boundary, is then a very desirable
property of a cosmological model.
A first result of the present paper is that
we determine the class of models with coupling different from
the standard choice $\phi^2 R$ which have asymptotic inflationary
attractors. We find that, in most models, either
the Universe does not inflate at all, or it does so toward the wrong
phase-space direction. The main result is that, in those cases in
which a successful inflation is recovered, the kinematics of
the Universe and the main
features of the fluctuation spectrum do not depend on $f(\phi)$.

\section{\normalsize\bf Inflationary attractors and the perturbation spectrum}
We start from the following Action [we assume Planck units, $G=c=\hbar=1$
and signature ($+---$)]
\beq\label{act2}
A=\vol\left[ -{R\over 16\pi}+\half\xi f(\phi)R+
\half g\umunu\phi\ddemu\phi\ddenu-V(\phi)\right].
\eeq
Great simplification is obtained passing to the the so-called
Einstein frame, i.e. deriving the field equation in the new metric
$\tilde g\dmunu=e^{2\ome}g\dmunu\,$ with
\beq\label{omega}
2\ome=\ln|1-\gam f|\,,
\eeq
where $\gam\equiv 8\pi\xi<0$.
We will work here assuming $\xi<0$, which guarantees that the new
metric is non-singular.
The old metric will be referred to as the Jordan frame.
The Einstein equations in the new frame are then
\beq\label{Eeinst}
\tilde G\dmunu= {1\over 1-\gam f}\left[
8\pi T\dmunu+{3\gam^2 f'^2\over 2(1-\gam f)} \left(
\phi\ddemu\phi\ddenu-
\half g\dmunu \phi\ddea\phi\udea\right)\right]\,,
\eeq
where $T\dmunu$ is the usual energy-momentum tensor for the
scalar field $\phi$. From now on, we confine ourselves to
 a homogeneous, isotropic and spatially flat metric with cosmic
factor $a(t)$ and Hubble function $H(t)=\dot a/a$.
Unless otherwise stated, all quantities are meant to be expressed
in the rescaled variables:  a dot denotes differentiation with respect
to the new time $t=\int e^{\ome(t')} dt'$, the Hubble function $H$ is in terms
of the new cosmic factor, and so on. If $K\dmunu=\phi\ddemu\phi\ddenu-
\half g\dmunu \phi\ddea\phi\udea$
 denotes the kinetic sector of $T\dmunu$ in the
original frame, the kinetic sector $\tilde K\dmunu$  in the new Einstein frame
can be written in the form
$\tilde K\dmunu=F^2(\phi)K\dmunu,$
where
\beq\label{kinfac}
F^2(\phi)=[16\pi(1-\gam f)+3\gam^2 f'^2]/16\pi(1-\gam f)^2\,.
\eeq
A canonical kinetic sector is then obtained
defining a new field
\beq\label{integr}
\psi=\int d\phi F(\phi)\,.
\eeq
This integral is not easily done, even with simple choices of
the coupling $f(\phi)$. However, a powerful simplification is
attained in the limit $|\phi|\to\infty$, the same limit in which
the effect of the coupling is greater and inflationary attractors
are found. Let us denote in the following with
a  subscript `{\it i}' a quantity evaluated at the initial time.
Then, if $f'^2$ grows faster than $f$, we can integrate
Eq. (\ref{integr}) into $c\psi=\ln [f(\phi)/f_i]\,,$
 and thus $f(\phi)=f_i\exp c\psi,$
where $c=(16\pi/3)^{1/2}$. It then follows that the initial value
of $\psi$ is $\psi_i=0$. When $f'^2$ grows
exactly like $f$, i.e. for $f=\phi^2$, one has
instead $c(\xi)=2[\gam/(6\xi-1)]^{1/2}$.

Suppose now that the following relation holds between the coupling $f$
and the potential: $V=\lam f^M\,.$
This  relation is verified,
for instance, if both  $V$ and $f$ are power functions of the
field, or if they are both exponentials.  The potential in the Einstein frame,
in the same limit as above, is then
\beq\label{pot2}
U(\psi)=\lam\gam^{-2} f^{\mu}=\beta \exp({c\mu \psi})\,,
\eeq
where $\beta\equiv
f_i^{\mu}\lam \gam^{-2}\,$, and $\mu\equiv M-2\,$.
We then explore the bidimensional parameter space $(\xi,M)$.
Remarkably, an asymptotic flat potential able to drive a slow-rolling
inflation in a FRW metric with scale factor $a(t)$
is found in a single case: $V\sim f^2$.
In all other cases, we find in the
$\dot\psi,\psi$ phase space the attractor
trajectories
\beq\label{attrpsi}
\dot\psi=A e^{c\mu \psi/2}\,,\quad A^2=2\beta \mu^2/ (9-\mu^2)\,,
\eeq
where the {\it negative} root for $A$ is to be chosen, corresponding to
$\phi$ decreasing toward its stable minimum, where the
Friedmann behavior takes place.
This corresponds to trajectories
\beq\label{attrphi}
f'\dot\phi=cAf_i^{-\mu/2}f^{(\mu+1)/2} \,,
\eeq
in the $(\dot\phi,\phi)$ plane.
In the
Appendix, solutions of this kind will be proven to be {\it attractors}; graphic
evidence of their attractive properties is provided by our Poincar\`e
projected phase spaces (Figs. 1-4). A power-law expansion (for $M\not=2$)
takes place on the attractors:
\beq\label{exactpl}
a=a_i\left(1+t/\tau\right)^p\,,\quad p=3/\mu^2\,
\eeq
(in the conformally rescaled time), where
$\tau=-2/Ac\mu>0 $. The cancellation of the coupling constant
$\xi$ from (\ref{exactpl}) occurs only in this class
of models: in the ordinary coupling $f=\phi^2$
the cancellation does not take place. Notice that the solution
(\ref{exactpl}) as well as the attractor (\ref{attrpsi}) above
are {\it exact} solutions as long as we take the potential (\ref{pot2}). From
(\ref{exactpl}) we see that $H=H_i(1+t/\tau)^{-1}$, where
$H_i=p/\tau$.

The behavior of the cosmic scale factor is accelerated, hence
inflationary, only if
$M<\sqrt{3}+2\sim 3.7$. However, for $M<2$ the model
enters an eternally inflating phase, since then $dU/d\psi<0$ and
the field $\psi$ is pushed to ever growing values,
never reaching the Friedmann phase located at $\psi\to-\infty$
[from the mapping $\psi\to\phi$ follows that $\psi\to-\infty$ when
$f(\phi)$ vanishes]. The only allowed range is then
\beq\label{narr}
2\le M< 3.7\,.
\eeq
Outside this narrow range, the model studied here does not allow
a successful chaotic inflation.
When $f=\phi^2$ the expansion is modified to $\hat p=p(1-1/2\xi)$.
This is inflationary for any $M$ if $\xi<0$ is small enough\bib{LS}.

In the original Jordan frame the cosmic scale expansion
is still a power-law in the Jordan time,
 but with an exponent $p_J=(p+\mu)/(1+\mu)$,
which is inflationary if $p>1$.

Since we have our theory in the Einsteinian form, we are
allowed to employ the standard formalism\bib{GPH} to derive the
perturbation spectrum at the horizon-crossing epoch.
We will see that we can determine the spectrum without
specifying the coupling function. We assume here that
the horizon-crossing condition is to be found
in the Einstein frame, i.e. that the condition only involves the
rescaled variables [see Ref. (\ref{GJ})].

The rms amplitude $\delta\rho/\rho$ of the perturbations
on a comoving scale $k^{-1}$ at horizon   crossing
(after the end of inflation) is  given by
\beq\label{del}
\delta\rho/\rho=\left(H^2/ \dot\psi\right)_{k=aH}\,,
\eeq
times an order-of-unity factor\bib{GJ}$^,$\bib{LM} depending on
the power-law exponent $p$. Along the solution (\ref{attrpsi}) we have
\beq\label{del2}
H^2/\dot\psi=(\tau/2) c\mu H_i^2(1+t/\tau)^{-1}
\eeq
Now, since
$k=aH=a_iH_i(1+t/\tau)^{p-1}\,,$
eliminating the time between the latter equation
and (\ref{del2}) we obtain
\beq\label{spe}
\delta\rho/\rho=
 (p/2)c\mu H_i(k/a_iH_i)^{1/(1-p)}\,,
\eeq
where the last factor expresses the scale dependence of the spectrum.
Since $p>1$, the spectrum grows with the scale, as it is
commonly found in power-law inflation, and becomes flat
for $p\to\infty$ (slow-rolling inflation). Our result
shows that this conclusion does not
depend on the choice of $f(\phi)$, as long as $V=\lam f^M$.

\section{\normalsize\bf Stochastic approach to fluctuations}
To compare the theory with the observed large-scale
structure we need also the higher-order
moments of the fluctuation distribution. In particular, it
is astronomically interesting to detect
possible deviations from Gaussianity of the primordial
spectrum\bib{QDOT}.
We investigate then the stochastic properties of adiabatic
fluctuations during a power-law inflation driven by the JBD field
$\phi$. The pioneering work in this context
is Ref. (\ref{STAR}).
Since we have now an Einsteinian theory with an exponential
potential\bib{LM}, we may analytically solve the Fokker-Planck equation
(FPE) for the probability distribution of the fluctuations.
The only other known cases of analytically solvable FPE are the
minimally coupled [$f(\phi)=0$] case with quartic or exponential
potentials\bib{OLM}.
In power-law inflation, contrary to slow-rolling inflation, the term
$\ddot\psi$ is not generally negligible in the field equation for $\psi$.
However, we show that its contribution is not dominant in the
cases investigated here, at least for $M$ not close to its upper
bound $3.7$.
The condition $|\ddot\psi | \ll |U'|$ gives in fact, when applied to the
attractor solution (\ref{attrpsi}),
$A^2 /2\beta\ll 1\, ,$
that is, $(c\mu )^2\ll 24\pi$.
We have, for instance,
$A^2/2\beta\sim 0.12$ for $M=3$, which is moderately satisfactory,
and  $A^2/2\beta\sim 0.03$ for $M=2.5$, a good approximation.
We can then safely put $\dot\psi=-U'/ 3H $
and  write the Langevin equation
in the useful time variable $\alp=\int Hd\tau$
\beqa\label{lan2}
{d\psi\over d\alp}=-{U'\over 3H^2}+\sqrt{3\over 8\pi}g H\eta(\alp)
=-{c\mu \over 8\pi}
+g\sqrt{\beta} e^{c\mu \psi/2}\eta(\alp),
\eeqa
where
$\eta(\alp)$ is the stochastic Gaussian noise due to the super-horizon
stretching of high frequency modes,
and $g$ is an order-of-unity constant\bib{LM} which depends on the
power-law exponent  $p$. We have made use in (\ref{lan2})
of  the ($0,0$) Einstein
equation in the Einsteinian frame in slow-rolling regime,
\beq
3H^2=8\pi\lam\gam^{-2}f_i^{\mu}\exp(c\mu\psi)\,.
\eeq
We will denote here with a subscript `{\it cl }' the
unperturbed quantities, i.e. the solution to the zero-noise
classical equations of motion, to distinguish them from
the stochastic variables which appear in the Langevin and
Fokker-Planck equations.
Let us define the new field variable
\beq\label{newf}
\Psi=\int {d\psi\over g\sqrt{\beta} e^{c\mu \psi/2}}=
-{2\gam\over c\mu g\sqrt{\lam}}f_i^{-\mu/2}e^{-c\mu\psi/2}\,,
\eeq from
which Eq. (\ref{lan2}) for $f\to\infty$ is simplified to
\beq\label{lan4}
d\Psi/ d\alp=\Psi/p+\eta(\alp)\,.
\eeq
The FPE associated to Eq. (\ref{lan4}) is in the Ornstein-Uhlenbeck\bib{RIS}
form
\beq\label{fpe2}
{\partial W\over \partial \alp}=-{1\over p}{\partial (\Psi W)\over
\partial\Psi}
+{\partial^2 W\over \partial \Psi^2}=-{\partial (D_1 W)\over \partial\Psi}
+{\partial^2 (D_{11}W)\over \partial \Psi^2}\,,
\eeq
where $D_1=\Psi/p$ is the drift coefficient and $D_{11}=1$ is the
diffusion coefficient.
One sees that the definition domain of $\Psi$ is ($0,+\infty$)
for $f(\phi)$ non-negative definite (remember that $\gam<0$).
Then we must impose
the reflecting boundary conditions
$S(\Psi=0)=0\,\,,S(\Psi\to\infty)=0$
where
\beq\label{current}
S=-D_1 W+ \partial (D_{11}W)/\partial \Psi\,,
\eeq
is the probability flux.
This introduces a `reflecting' term in the solution\bib{LM}, which
turns out to be negligible at large times, when
$<\Psi>$ moves to infinity.
The general solution\bib{RIS} of (\ref{fpe2}) with initial condition
$W(\Psi,0)=\delta(\Psi-\Psi_i)$ is then found to be
\beq\label{fpsol2}
W(\Psi,\alp)=
\sqrt{1\over 2\pi p \left(e^{2\alp/p}-1\right)}\left\{
\exp\left[- {(\Psi-\Psi_{cl})^2\over
2p(e^{2\alp/p}-1)}\right]+
\exp\left[ -{(\Psi+\Psi_{cl})^2\over
2p(e^{2\alp/p}-1)}\right]\right\},
\eeq
where $\Psi_{cl}=\Psi_i e^{\alp/p}$.
It is easy to see that all the moments of $\psi$
are defined. In fact, the integral
\beq\label{kth}
M_{\kappa}(\psi)\equiv
<\psi^{\kappa}>=\int_{-\infty}^{\infty}\psi^{\kappa}W(\psi,\alpha)
d\psi\,,
\eeq
where $W(\psi,\alpha)=W(\Psi,\alpha)|J|$, and
$|J|=d\Psi/ d\psi$,
converges for any $\kappa$. The same is not always true for the moments of the
original field $\phi$. For instance, if $f(\phi)=\phi^{2m}$,
we have that  all the moments $<\phi^{\kappa}>$ for which
$\kappa\ge m\mu$ are divergent for $\phi\to\infty$.

The distribution of the field $\psi$ is non-Gaussian. We are interested
in estimating its deviation from Gaussianity through the
skewness coefficient
\beq\label{skk}
s=M_3(\psi-\psi_{cl})/M_2^{3/2}(\psi-\psi_{cl})\,.
\eeq
In the following
we will neglect the contribution from the reflecting term,
so that we have $M_1(\psi)=\psi_{cl}$.
We may first calculate approximately the second central
moment of $\psi$, i.e. the variance, in the limit of
small $|\psi-\psi_{cl}|$.
We will see that
this is completely justified in viable theories of inflation,
due to the smallness of $H$, which sets the amplitude
of the diffusion around the classical trajectory. Expanding
$|\Psi-\Psi\cl|$ in powers of $|\psi-\psi\cl|$ we
can write the solution $W(\psi)$ to the first order
as a Gaussian with mean $\psi\cl$ and variance
\beq\label{var}
D=pg^2(e^{2\alp/p}-1)[\lam\gam^{-2}f_i^{\mu}
e^{c\mu\psi_{cl}}]
\,.\eeq
We can now recognize in the square brackets, apart order-of-unity
factors, the classical value of $H^2$. It follows that,
for large $\alp$,
\beq
\sqrt{<(\psi-\psi_{cl})^2>}\approx H\cl e^{\alp/p}=H\cl (t/\tau)
\,.\eeq
For $t\gg \tau$ we have $H\cl t/\tau\sim H_i$.
In order to avoid
exceedingly large perturbations on the microwave sky, $H_i$ must
be smaller than, roughly, $10^{-4}$ in Planck units, as it is
evident from Eq. (\ref{spe}). It turns out that the deviation
of $\psi$ from $\psi_{cl}$ is indeed very small,
which justifies the approximations made above.
We may now calculate the third central moment $M_3(\psi-\psi_{cl})$
in the following way. Let us consider a Gaussian distribution $W(\Psi)$
normalized to unity,
with mean $\Psi_{cl}$ and variance $\sigma$.
We transform it under the mapping $\Psi=\Psi(\psi)$ to $W[\Psi(\psi)]
|\Psi'|$, where $\Psi'=d\Psi/d\psi$.
The inverse mapping is $\psi=\psi(\Psi)$.
The mean of $\psi$ is now
$\psi_{cl}=\psi(\Psi_{cl})$, while the variance and the higher-order moments
have to be calculated.
The new distribution is $W(\psi)=W[\Psi(\psi)]|\Psi'|$.
In the limit $|\psi-\psi_{cl}|\to 0$ we may expand $\Psi(\psi)$
and $\Psi'$ in Taylor series around $\psi_{cl}$.
It follows
\beq\label{newg2}
W(\psi)|\Psi'|=(2\pi D)^{-1/2}\exp
\{-{1\over 2D}\left[(\psi-\psi_{cl})^2-R(\psi-\psi_{cl})\right]\}
\,,\eeq
where $D=\sigma |\Psi'_{cl}|^{-2}$ and
$R=2D\Psi''_{cl}/\Psi'_{cl}$.
Here $\Psi'_{cl},\Psi''\cl$ mean $\Psi'$ and $\Psi''$,
respectively, calculated in $\psi=\psi_{cl}$.
The distribution (\ref{newg2}) is again
a Gaussian curve, but with mean, variance and normalization
different from the original ones.
After some simple manipulations it is easy to find
\beq\label{easy}
M_3(\psi-\psi_{cl})={R\over 2}\left(3D+{R^2\over4}\right)\,.
\eeq
Let us specialize to our model. We have then
that $D$ is given in (\ref{var}), while $R=-c\mu D$. In the limit
of small $D$, i.e. small $H^2$, we have
\beq\label{skew}
s=-3c\mu D^{1/2}\,/2\sim -H\cl(t/\tau)
\,.\eeq
The deviation from a Gaussian distribution turns out
to be indeed negligible as long as the initial value for $H$ is
less than $10^{-4}$ in Planck units. In other words,
the need for a smooth microwave background has the consequence
of a very small deviation from Gaussianity. Perturbations which crossed out
the horizon when $H\sim 1$ can  have a very skewed distribution, but they are
now much larger than the observable Universe.
Similar conclusions
hold for the higher moments of the distribution.

\section{\normalsize\bf Conclusions}
One of the most appealing aspects of chaotic inflation is that even
the simplest models possess the required properties for
a successful inflation. Almost all the possible initial conditions
for a scalar field dominated Universe with a simple potential lead
to a conspicuos period of inflation. This feature implies,
in the dynamical
phase-space of such theories, the existence of attractor trajectories
over which the other trajectories converge.
It turns out, contrary
to the uncoupled case, $f=0$, and to the usual $f=\phi^2$ case, that models
with generic coupling satisfy the chaotic inflation requirements
only when $2\le M< 3.7$.
The resulting power-law expansion is independent of $f(\phi)$.
This power-law case is noticeable because
it can be adopted as a chaotic hyperextended inflation with a
kinematics independent of the coupling function.

The scale-dependence, $k^{1/(1-p)}$, also independent of $f(\phi)$,
is an interesting
signature of power-law inflation, possibly testable with the
data from COBE. Solving the Fokker-Planck equation we
can evaluate higher-order moments of the fluctuation
distribution. This shows that the deviation from Gaussianity on
observable scales is negligible in all acceptable inflationary models of the
kind studied here.

Finally, we prove analytically in the Appendix the stability of the
attractors.

\vspace{24pt}
\centerline{\bf Appendix }
Numerical phase-spaces show with full evidence the stability
of the trajectories labelled as $A$-type in the text. Here we prove
it analytically for the case $M>2$. As it was shown in Sect. 2,
the phase-space in the large $\phi$ limit is governed, in
the rescaled variables,  by the dynamical system
($x=\psi, y=\dot \psi$)
\beq\label{Asys}
\dot x = y\,,\qquad \dot y = -3H(x,y)y -U'(x)\,,
\eeq
where $U'=\beta c\mu  e^{c\mu x}$. There is an attractor trajectory
\beq\label{Aattr}
y_0=Ae^{c\mu x/2}\,,
\eeq
with $c\mu >0$ and $A<0$.
Here we assume for simplicity that $H$ is
calculated along the attractor $H=[8\pi(1/2+\beta/A^2)/3]^{1/2}|y_0|$,
freezing the $x,y$-dependence of $H$ on the assumed solution (\ref{Aattr}).
Dealing with the full function $H=H(x,y)$ only entangles the algebraic details
without changing the conclusions.
To show that (\ref{Aattr}) is indeed an attractor, let us define a new set of
coordinates $u,v$, such that $u=0$ defines the trajectory (\ref{Aattr}):
$u=y-Ae^{c\mu x/2},\quad v=y+Ae^{c\mu x/2}\,.$ Under this mapping,
the dynamical system (\ref{Asys}) becomes
\beqa\label{Anewsys}
2\dot u&=& {c\mu \over 4}(u^2-v^2)-3H(u+v)-2U'\nonumber\,,\\
2\dot v &=&{c\mu \over 4}(v^2-u^2)-3H(u+v)-2U'\,.
\eeqa
We now linearize (\ref{Anewsys})
around $x_0, y_0=Ae^{c\mu x_0/2}$, \ie around a generic point on
$Ae^{c\mu x/2}$. This corresponds to linearizing around
$u=0, v=v_0=2Ae^{c\mu x_0}$. Notice that $v_0<0$.
We get then for the coordinate $u$ the following equation
(the equation for $v$ is not of concern in this
context)
\beq\label{Alinsys}
\dot u = a_1 v +a_2 u\,.
\eeq
It turns out that $a_1=0$: this  confirms that the trajectory
$u=0$ is a solution
of the system (\ref{Asys}). The sign of $a_2$ determines the stability
of the attractor: when $a_2$ is negative, in
fact, $u\to 0$ starting from any small
perturbation $u_0$.
The process evolves with a time-scale $1/|a_2|$. We have
\beq\label{a_2}
a_2=v_0 c\mu [(18-\mu^2)/ 2\mu^2]
\eeq
which is indeed negative because $v_0$ is negative and $m<\sqrt{3}$.
The stability time scale $1/|a_2|$ goes like $(\beta)^{-1/2}$,
i.e. like $\xi\lam^{-1/2}$: the process of infalling toward the
attractor is faster as $\xi\to 0$ or $\lam\to\infty$.


\vspace{.5in}
\newpage
\centerline{\bf REFERENCES}
\footnotesize
\frenchspacing
\begin{enumerate}

\item\label{JBD}
	P. A. M. Dirac, Proc. R. Soc. {\bf A338}, 439  (1974) ;
	 P. Jordan, {\it Schwerkraft und Weltall} (Wieweg, Braunschweig, 1955);
	C. Brans and R. H. Dicke \rpr, {\bf 124}, 925 (1961).
\item\label{BD}
N. D. Birrel and P. C. W. Davies, {\it Quantum Fields in
Curved Spaces} (Cambridge Univ. Press, Cambridge, 1982).

\item\label{SS}
S. Randjbar-Daemi, A. Salam, and J. Strathdee, \rpl {\bf B135}, 388
(1984); K. Maeda, \rcqg {\bf 3}, 233 (1986);
T. Applequist and A. Chodos, \rprl {\bf 50}, 141 (1983).
\item\label{ZA}
	A. Zee, \rprl, {\bf 42}, 417 (1979);
	F. S. Accetta, D. J. Zoller and M. S. Turner, \rpr D{\bf 31},
	3046 (1985).
\item\label{LS}
	D.  La and P. J. Steinhardt, \prl, {\bf 62}, 376 (1989).
\item\label{AT}
	D. La and P. J. Steinhardt, \rpl {\bf B220}, 375 (1989);
	F. S. Accetta and J. J. Trester, \rpr D{\bf 39}, 2854 (1989);
        R. Holman, E. W. Kolb, and
	Y. Wang, \rprl {\bf 65} 17 (1990);
	R. Holman, E. W. Kolb, S. L. Vadas and Y. Wang, CMU-HEP 90-09,
	FNAL Pub-90/99-A (1990) preprint; A. Linde, CERN-TH-5806/90 preprint
	(1990); Y. Wang, \rpr D{\bf 42}, 2541 (1990);
L. Amendola, S. Capozziello, M. Litterio and F. Occhionero, \rpr {\bf
D45}, 417 (1992).

\item\label{KFM}
	U.  Kasper, Nuovo Cim. {\bf B103}, 291 (1989);
	T. Futamase and K. Maeda,\pr D{\bf 39}, 399 (1989);
	L. Amendola, M. Litterio and F. Occhionero,
	\ijmp A, {\bf 5}, 3861 (1990).
\item\label{SA}
	P. J. Steinhardt and F. S. Accetta,
	\rprl {\bf 64},  2740 (1990).
\item\label{AL}
L. Amendola, preprint (1993), to be published in \rpl {\bf B}.
\item\label{DR}
R. De Ritis, talk delivered at the ``X Italian Conference on
General Relativity and Gravitational Physics", Bardonecchia, Sept. 1-5
1992.
\item\label{GJ}
A. H. Guth and B. Jain, \pr {\bf D45}, 426 (1992);
D. H. Lyth and E. D. Stewart, \pl B{\bf 274}, 168 (1992).
\item\label{LIN}
	A. Linde, \rpl {\bf 129B}, 177 (1983);
	 \rnp {\bf B216}, 421 (1983).
\item\label{W}
	B. Whitt, \pl {\bf 145B}, 176 (1984);
	J. D. Barrow and Cotsakis S.,\pl, {\bf 214B}, 515 (1988);
	K.-I. Maeda, \pr D{\bf 39}, 3159 (1989);
	H.-J. Schmidt,{\it Class. Q. Grav.} {\bf 7}, 1023 (1990).
\item\label{LM}
	F. Lucchin and S. Matarrese, \pr D{\bf 32}, 1316 (1985).
\item\label{OLM}
	A. Ortolan, F. Lucchin and S. Matarrese, \pr D{\bf 38}, 465 (1988).
\item\label{QDOT}
W. Saunders \etal, \nat {\bf 349},32 (1991).
\item\label{STAR}
	A. A. Starobinsky, in {\it Field Theory, Quantum Gravity and
	Strings}, eds. H.J. de Vega and N. Sanchez (Springer-Verlag,
	Lect. Notes in Physics 246, 1986)
\item\label{RIS}
	H. Risken, {\it The Fokker-Planck Equation} (Springer-Verlag,
	Berlin, 1989).

\item\label{GPH}
A. Guth and S.-Y. Pi, \prl {\bf 49}, 1110 (1982);
S. W. Hawking, \pl B {\bf 115}, 339 (1982);
A. A. Starobinsky, \pl B {117}, 175 (1982).

\item\label{COBE}
G. F. Smoot \etal, COBE preprint 92-02 (1992).
\end{enumerate}

\vspace{.3in}

\newpage
\centerline{\bf FIGURE CAPTIONS}
\begin{enumerate}

\item\label{f1}  The phase-space portrait of the model with
$f(\phi)=\phi^{2m}$ and $V(\phi)\sim\phi^{2n}$, where $n=1, m=2$ and
negative $\xi$ is displayed. Notice the  two outward directed attractors
and the central Friedmann region. This model is outside
the successful range of parameters.
All the trajectories
spring out from the singular points at infinity on the Poincar\'e
circle at angles $\pi/2$ and $3\pi/2$. The diagram is symmetric
with respect to the origin.

\item\label{f2}
As in Fig. 1, with $n=5, m=2$. This model is inside
the successful range: the power-law attractors have a natural end.
\item\label{f3}
As in Fig. 1, with $n=4, m=2$. This model satisfies the condition $M=2$.
Along the deSitter-like attractors one has $\ddot\phi=\dot\phi\to 0$.
\item\label{f4}
As in Fig. 1, with $n=0, m=1$. This phase-space is equivalent to
extended inflation.

\end{enumerate}

\end{document}